\documentclass[preprint,12pt]{aastex}
\def\be{\begin{equation}}
\def\ee{\end{equation}}

\def\vhel{$V_{\rm hel}$}

\begin{document}
\title{Structure of Supergiant Shells in the Large Magellanic Cloud}
%
%
\author{Laura G.\ Book\altaffilmark{1,2,3}, You-Hua Chu\altaffilmark{2}, Robert 
A.\ Gruendl\altaffilmark{2}}
\altaffiltext{1}{\itshape Department of Physics,
 University of Illinois at Urbana-Champaign, 1110 West Green Street, Urbana, 
IL 61801}
\altaffiltext{2}{\itshape Department of Astronomy, University of Illinois 
at Urbana-Champaign, 1002 West Green Street, Urbana, IL 61801}
\altaffiltext{3}{\itshape Current address: California
Institute of Technology, MC130-33, 1200 E. California Blvd., Pasadena, CA 91125}
%
%

%
%
%
\begin{abstract}
Nine supergiant shells (SGSs) have been identified in the Large 
Magellanic Cloud (LMC) based on H$\alpha$ images, and twenty-three SGSs 
have been reported based on \ion{H}{1} 21-cm line observations, but these 
sets do not always identify the same structures. We have 
examined the physical structure of the optically identified SGSs using 
\ion{H}{1} channel maps and P-V diagrams to analyze the gas kinematics. 
There is good evidence for seven of the nine optically identified SGSs to 
be true shells. Of these seven H$\alpha$ SGSs, four are the ionized inner
walls of \ion{H}{1} SGSs, while three are an ionized portion of a larger
and more complex \ion{H}{1} structure. All of the H$\alpha$ SGSs are
identified as such because they have OB associations along the periphery 
or in the center, with younger OB associations more often found along the 
periphery. After roughly 12 Myrs, if no new OB associations have been 
formed a SGS will cease to be identifiable at visible wavelengths. Thus, 
the presence and location of ionizing sources is the main distinction between 
shells seen only in \ion{H}{1} and those also seen in H$\alpha$. Based on
our analysis, H$\alpha$ observations alone cannot unambiguously identify 
SGSs, especially in distant galaxies.

\end{abstract}
%
%
%
\section{Introduction}  \label{sec:intro}

Observations of the interstellar medium (ISM) of spiral and irregular
galaxies have revealed complex filamentary structures indicative of a
violent ISM \citep{MS79}. The largest of these structures are the
supergiant shells (SGSs), with diameters approaching 1 kpc \citep{gm}.
SGSs are thought to be formed by the fast stellar winds and supernova
explosions of multiple OB associations; they require $10^{52}$--$10^{53}$ 
ergs for their creation, the equivalent of tens to hundreds of supernova
explosions \citep{M80}. The diameter of these shells often exceeds the 
scale height of the galactic gas disk, allowing them to puncture the 
gas disk and vent their hot interior gas into the galactic halo.  
The expansion of SGSs may also cause further star formation.  It has 
been suggested that the compression of the ISM on the rims of these 
shells and its subsequent gravitational collapse may be a significant 
cause of self-propagating star formation in galactic disks, thereby 
providing important insight into galactic evolution.  To examine the 
impact of SGSs on their host galaxy, it is necessary to determine the 
physical structure of SGSs that have been identified based on their
morphology alone, since some may be merely chance superpositions of
unrelated filaments.

The Large Magellanic Cloud (LMC) is an excellent site to explore the
nature of SGSs.  With its low inclination there is little line-of-sight
confusion, and its proximity (50 kpc; Feast 1999) allows us to probe
its ISM in a detail impossible for most other galaxies. SGSs in the LMC
were first identified by \citet{gm}, based on visual inspection of
H$\alpha$ images of the ionized ISM. This original paper tabulated four
SGSs, a number that was later expanded to nine by \citet{M80}. Later work
on LMC SGSs, however, has also been done based on observations of
\ion{H}{1}, the neutral atomic ISM. \citet{Kim99} examined the LMC in
\ion{H}{1} and tabulated 23 SGSs that have diameters greater than 360 pc.
Curiously, there is not a one-to-one correspondence between the largest
\ion{H}{1}-selected SGSs and the nine H$\alpha$-selected SGSs.

To investigate the physical structure of the H$\alpha$-selected SGSs 
and their relationship with the \ion{H}{1}-selected SGSs, we have 
examined the \ion{H}{1} environment of the H$\alpha$-selected SGSs 
using the H$\alpha$ images from the Magellanic Cloud Emission Line 
Survey \citep[MCELS;][]{MCELS05} and the \ion{H}{1} synthesis maps 
made with combined observations from the Australia Telescope 
Compact Array and Parkes Telescope \citep{Kim03}.  In this paper we 
describe these data sets in \S 2 and our methodology in \S 3, and 
critically examine the physical structure of the nine H$\alpha$-selected
SGSs in \S 4.  We discuss the nature of the SGSs and compare the
\ion{H}{1}-selected and  H$\alpha$-selected shells in \S 5.  A summary
is given in \S 6.

\section{H$\alpha$ and \ion{H}{1} Observations}  \label{sec:obs}

The MCELS H$\alpha$ images were taken with a CCD camera on the Curtis 
Schmidt Telescope at Cerro Tololo Inter-American Observatory (CTIO). 
The images have an angular resolution of $\sim$3$''$, and were taken 
with a narrow-band interference filter centered on the H$\alpha$ line 
($\lambda_{c} = 6563$ \AA, $\Delta\lambda=30$ \AA). The detector was 
a front-illuminated STIS 2048 $\times$ 2048 CCD with 21 $\mu$m pixels, 
giving a field of view of 1\fdg1 $\times$ 1\fdg1 \citep{MCELS}. The 
H$\alpha$ images were mosaicked to cover the central 8$^{\circ}$ 
$\times$ 8$^{\circ}$ of the LMC.  From the mosaic we extracted regions 
covering the neighborhoods of the nine H$\alpha$-selected shells of
\citet{M80}.

The neutral hydrogen data were obtained by \citet{Kim03} using the 
Australia Telescope Compact Array (ATCA) and combining with single-dish
observations using the Parkes Telescope. These observations cover 11\fdg1 
$\times$ 12\fdg4 of the sky, and have an angular resolution of 1\farcm0 and 
a resulting pixel size of $20''$. The observing band was centered on 1.419 
GHz, corresponding to a central heliocentric velocity of 297 km~s$^{-1}$. 
The bandwidth used was 4 MHz, giving a complete velocity range of $-$33 to 
+627 km~s$^{-1}$ \citep{Kim98,Kim03}.  Heliocentric velocities 
($V_{\rm hel}$) are used throughout this paper.

\section{Methodology}  \label{sec:method}

The first step in our analysis of the optically identified SGSs was to 
locate the nine shells identified by \citet{M80} within the H$\alpha$ 
image of the LMC. Using the rough sketch provided by \citet{M80} as a 
guide, we found structures which resembled their representation on the 
drawing. Some of these structures were difficult to locate, such as SGS 
7, 8 and 9, because they are faint, incomplete and irregular. For each 
SGS we then extracted regions with roughly twice the linear size of 
the optical shell from the H$\alpha$-mosaic and the \ion{H}{1} data cube,
with the exception of SGS 4 and 5 which were grouped together.

The \ion{H}{1} data cube provides both column density and velocity 
information along each line of sight. We began our investigation into 
the structure of a SGS by examining its \ion{H}{1} column density map, 
which is integrated over all velocities. We then examined the kinematics 
of the gas using a series of isovelocity images, or channel maps, to 
show the distribution of gas at different velocities. The channel maps 
typically spanned a range of 100 km~s$^{-1}$, at intervals of 4 - 5 km 
s$^{-1}$. We used the channel maps to estimate the systemic velocity
of a shell and used the map for the systemic velocity to determine 
whether the structure has a central cavity and whether the rim has 
brighter emission than the surroundings, indicating a shell structure 
rather than a hole.

Once the general kinematic structure of a shell was ascertained from 
the channel maps, we used this information to guide the selection of 
positions for position-velocity (P-V) diagrams. The P-V diagrams are 
produced by collapsing a narrow strip of the \ion{H}{1} data cube along 
its short axis to show the distribution of velocities along the longer 
axis of the strip, equivalent to an optical long-slit spectrum. These P-V 
diagrams are ideal for quantitative expansion velocity measurements, and 
also tell us whether the shell has a blowout structure.

Images of these data were created and overlaid using the plotting 
task CGDISP in MIRIAD, a radio interferometry data reduction package, 
and WIP, a graphics software package, both created for the 
Berkeley-Illinois-Maryland Association (BIMA) Radio array. Analysis 
of the data was also done using the astronomical data processing package 
SAOimage DS9, developed by Smithsonian Astrophysical Observatory, which 
provided a user-friendly interface for explorations of datasets.  

\section{Analysis of Individual SGSs} \label{sec:SGS}

The nine SGSs cataloged by \citet{M80} were identified based on a
shell-like structure in H$\alpha$, and they have been commonly
referred to as LMC 1-9.  To distinguish this sample from the 23
\ion{H}{1}-selected shells \citep{Kim99}, we will call the latter 
SGS 1-23.  The \ion{H}{2} region names cited in this section are
``N'' from \citet{hen} and ``DEM L'' from \citet{dem}.  The ``LH''
designation of OB associations is from \citet{lh}.

\subsection{LMC 1}  \label{sec:lmc1}

\begin{figure}[!ht]
\centering
\includegraphics[height=3in]{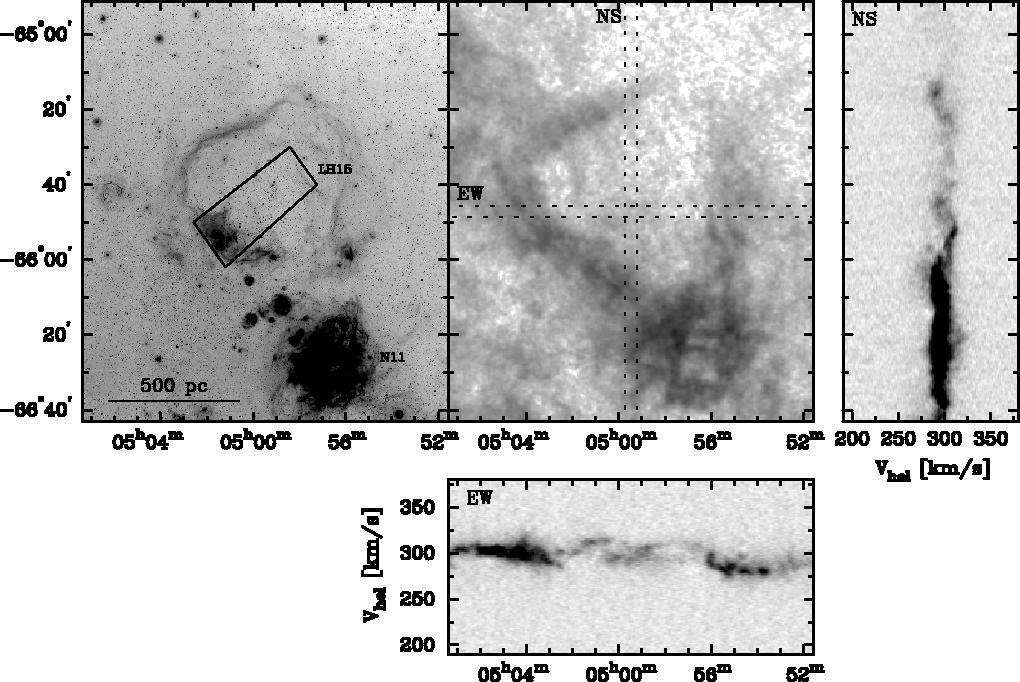}
\caption{An H$\alpha$ image of supergiant shell LMC 1(left panel), 
marked with relevant OB associations and \ion{H}{2} regions. An \ion{H}{1}
column density map (center panel), and P-V diagrams (right and bottom panels)
whose positions are overlaid on the \ion{H}{1} map.} 
\label{fig:SGS1a}
\end{figure}

The SGS LMC 1 is located in the northwest corner of the LMC, with
central coordinates 5$^{\rm h}$00$^{\rm m}$, $-$65$^{\circ}$40$'$ (J2000). 
The optical shell of LMC 1 has a diameter of $\sim$~50$'$, corresponding 
to 750~pc (see Fig.~\ref{fig:SGS1a}). The H$\alpha$ image shows that the 
southeast and northeast edges of the shell are brighter than the western
side, indicating a density gradient of ionized hydrogen from east to west
over the shell. The \ion{H}{1} shell, which is larger than the H$\alpha$ 
shell and lies outside of it, also shows this density gradient, with the
southeast wall much brighter than the other sides.  The \ion{H}{1} column
densities at the northwest corner of LMC 1 are so low that the gray-scale
image appears to show a gap.

\begin{figure}[!ht]
\centering
\includegraphics[height=3in]{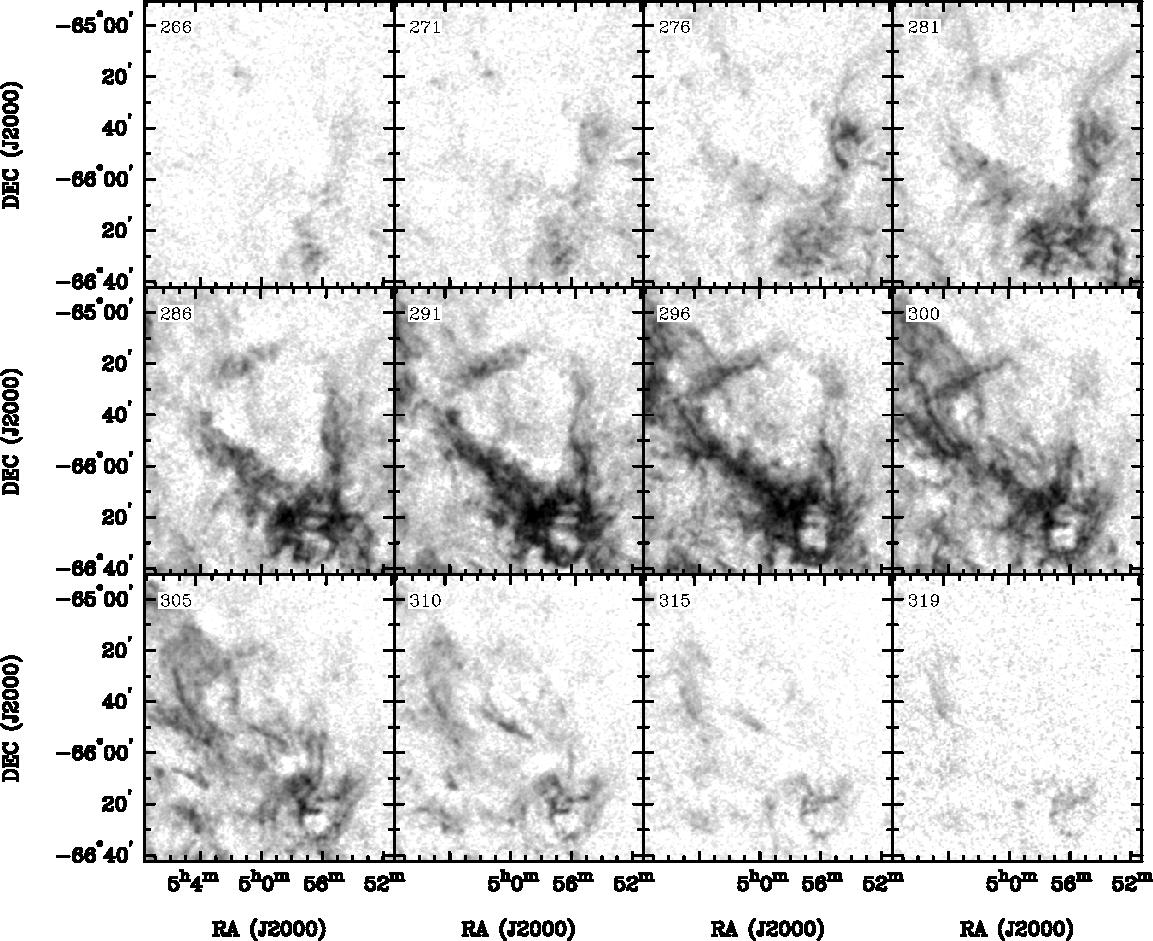}
\caption{\ion{H}{1} channel maps of supergiant shell LMC 1 , whose field of 
view is identical to that of the H$\alpha$ and \ion{H}{1}
images in figure \ref{fig:SGS1a}, but with a slightly different coordinate projection.} 
\label{fig:SGS1b}
\end{figure}

LMC 1 is detected in channel maps (Fig.~\ref{fig:SGS1b}) in the velocity 
range of 265 to 315~km s$^{-1}$ with a systemic velocity of 
$\sim$~295~km~s$^{-1}$.  The channel maps near the systemic
velocity show that the rim of LMC 1 is complete with no gap in the 
northwest corner.  The north and west sides of the shell appear to be 
blueshifted by up to 20~km~s$^{-1}$ relative to the systemic velocity, 
while the south and east sides are redshifted by a similar amount. 
In the 310--315~km~s$^{-1}$ channel maps, a bright \ion{H}{1} filament
is visible within the shell interior, and this filament is spatially
coincident with the surface of the bright ionized southeast rim of
LMC 1.  The channel maps also show a small shell to the southwest 
of LMC 1, corresponding to the \ion{H}{2} complex N11.

To determine the expansion velocity and kinematic structure of LMC 1, 
we use P-V diagrams along the north-south and east-west cuts passing 
through the shell center (see Fig.~\ref{fig:SGS1a}). In the north-south P-V 
diagram, we see irregular expansion with a maximum velocity offset of 
15~km~s$^{-1}$ on the receding side of LMC 1, but detect no counterpart 
on the approaching side. The east-west P-V diagram shows a more 
complex velocity structure. It shows a similar irregular expansion on 
the receding side, but within the central region there is an 
additional broad component near the systemic velocity.  The spatial 
distribution of this turbulent, low-velocity material can be seen in 
the channel maps near the systemic velocity.

LMC 1 appears to be a real, physical shell, corresponding to the ionized 
inner wall of the \ion{H}{1} shell SGS 3. The \ion{H}{1} shell appears 
to be one-sided, with irregular expansion of $\sim$15~km~s$^{-1}$. The 
OB association LH15 stretches from the southeast rim of LMC 1 northeast 
into its interior, and is responsible for the ionization of the shell. 
The supernovae and fast stellar winds from this OB association are most
likely responsible for the kinematic structure of the \ion{H}{1} shell.
A blow-out in the approaching side of LMC 1 must have occurred, as 
indicated by a lack of \ion{H}{1} at blueshifted velocities within the 
central cavity.  Blow-outs prevent the build-up of hot gas
and thermal pressure in shell interiors.  Without a hot pressurized
surrounding, supernova remnant shocks are less likely to be thermalized;
instead, they may advance outwards, impact the shell walls, and 
accelerate interstellar gas only locally.  This explains LMC 1's 
irregular expansion and accelerated material at peculiar velocities.

\subsection{LMC 2}  \label{sec:lmc2}

\begin{figure}[!ht]
\centering
\includegraphics[height=4in]{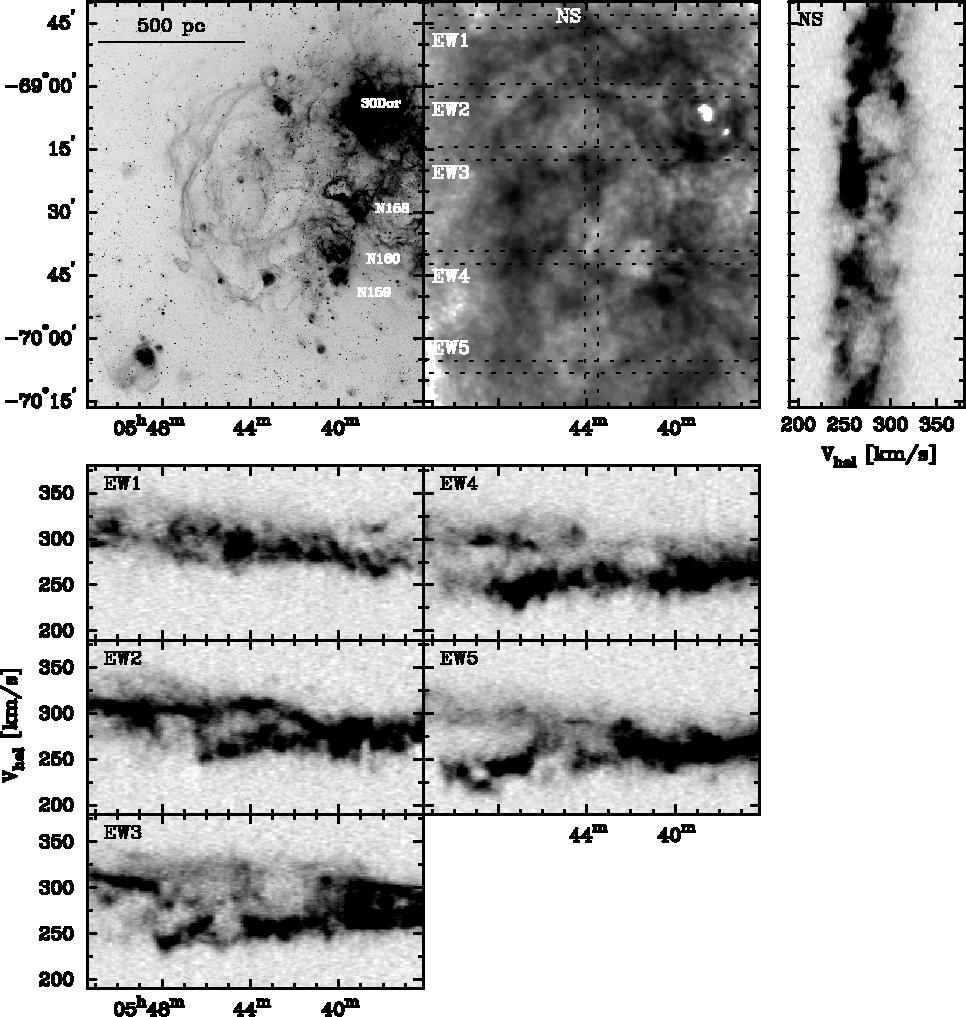}
\caption{Same as figure \ref{fig:SGS1a}, but Supergiant Shell LMC 2.} 
\label{fig:SGS2a}
\end{figure}

LMC 2 is a large structure on the southeast of the 30 Doradus Nebula, 
with central coordinates 5$^{\rm h}$44$^{\rm m}$, $-$69$^{\circ}$20$'$ 
(J2000) and a diameter of 900 pc (see Fig.~\ref{fig:SGS2a}). It is the 
brightest SGS in the LMC in H$\alpha$.  The shell consists of a set 
of long H$\alpha$ filaments along its northern, eastern and southern 
edges, and a ridge of active star formation regions on its western 
edge, including N158, N159, and N160. Many studies have been made on 
the structure and kinematics of LMC 2, some finding the existence 
of a three-dimensional expanding structure \citep{C82} and others 
concluding that the ionized gas is confined between two sheets of 
\ion{H}{1} gas and expands only eastward \citep{P99}. Diffuse X-ray 
emission has been detected from the interior of LMC 2, based on which 
\citet{WH91} concluded that the shell was blowing out of the disk.

\begin{figure}[!ht]
\centering
\includegraphics[height=3in]{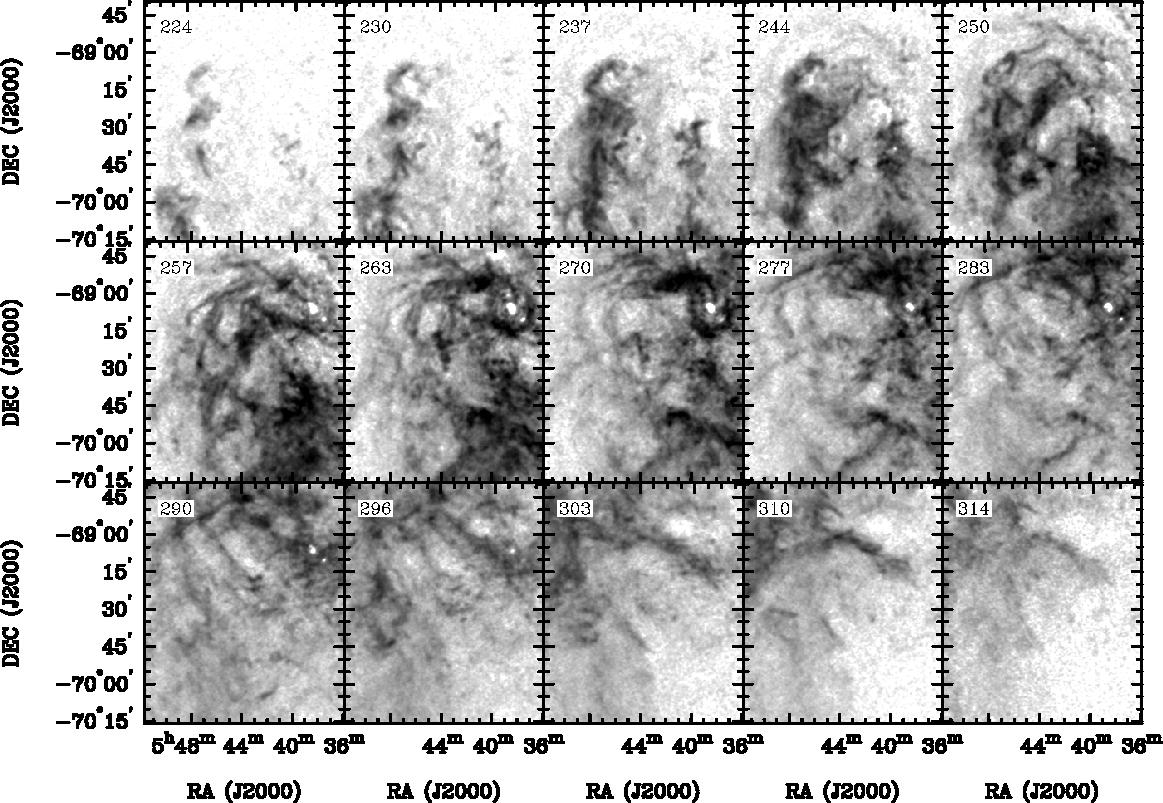}
\caption{Same as figure \ref{fig:SGS1b}, but Supergiant Shell LMC 2.} 
\label{fig:SGS2b}
\end{figure}

The \ion{H}{1} column density map shows a very complex structure, both 
within LMC 2 and in the surrounding region. There is significant 
structure interior to LMC 2 both in H$\alpha$ and \ion{H}{1} images. 
Upon closer examination, \ion{H}{1} enhancements often coincide with 
H$\alpha$ enhancements in the form of filaments, \ion{H}{2} regions, 
or diffuse emission. The \ion{H}{1} image does not show a contiguous 
cavity throughout LMC 2, but rather several localized cavities. The 
channel maps confirm the general correspondence between \ion{H}{1} 
enhancements and H$\alpha$ filaments and the lack of a 
single central cavity within LMC 2 \citep[figure \ref{fig:SGS2b}][]{P99}.

The physical structure of LMC 2 is best seen in P-V diagrams. The 
north-south P-V diagram passing through the center of LMC 2 
(see Fig.~\ref{fig:SGS2a} NS) shows an expanding shell consisting of 
the northern two-thirds of the optical shell, the southern boundary 
of which corresponds to a prominent curved east-west H$\alpha$ 
filament. This expanding shell structure is justified because the 
\ion{H}{1} gas converges to a single velocity component to the north
and south, and the east-west P-V diagram through this region (see 
Fig.~\ref{fig:SGS2a} EW2 and EW3) shows expansion and converges back to
a single velocity component to the west. The southern third of LMC 2 
shows a more complex structure, with a localized expanding cavity on 
the west (see Fig.~\ref{fig:SGS2a} EW4 and EW5). To the south of LMC 2, 
the P-V diagrams still show an expanding shell structure 
(see Fig.~\ref{fig:SGS2a} EW5 and NS), corresponding to a cavity in 
the \ion{H}{1} column density map. It is not clear whether this 
cavity is an extension of LMC 2, since we do not see clear evidence 
for a wall between them.
  
We have made 20 east-west P-V diagrams in 5$'$ steps over LMC 2 and 
the surrounding region. Because of the large area covered, 
$\sim$~76$'\times$95$'$, we are able to see that the two \ion{H}{1} 
sheets reported by \cite{P99} are part of an expanding structure. To 
the west, the two velocity components converge to 275 km~s$^{-1}$; 
to the east, they appear to diverge over the area of LMC 2 and beyond, 
but converge further to the east (see Fig.~\ref{fig:SGS2a} EW2, EW3, 
EW4 and EW5).  

We conclude that the optical SGS LMC 2 is not a simple expanding 
shell, but part of a more extended, complex \ion{H}{1} expanding 
structure. On a large scale, the \ion {H}{1} shows expansion over a 
region 1150 pc~(N-S)~$\times$ 850 pc (E-W). Within this overall 
structure there are substructures, such as the two expanding 
structures we identified in the northern two-thirds of LMC 2 
and to the south, which roughly correspond to the \ion{H}{1} SGS 
19 and 20, respectively. 

\subsection{LMC 3}  \label{sec:lmc3}

\begin{figure}[!ht]
\centering
\includegraphics[height=3in]{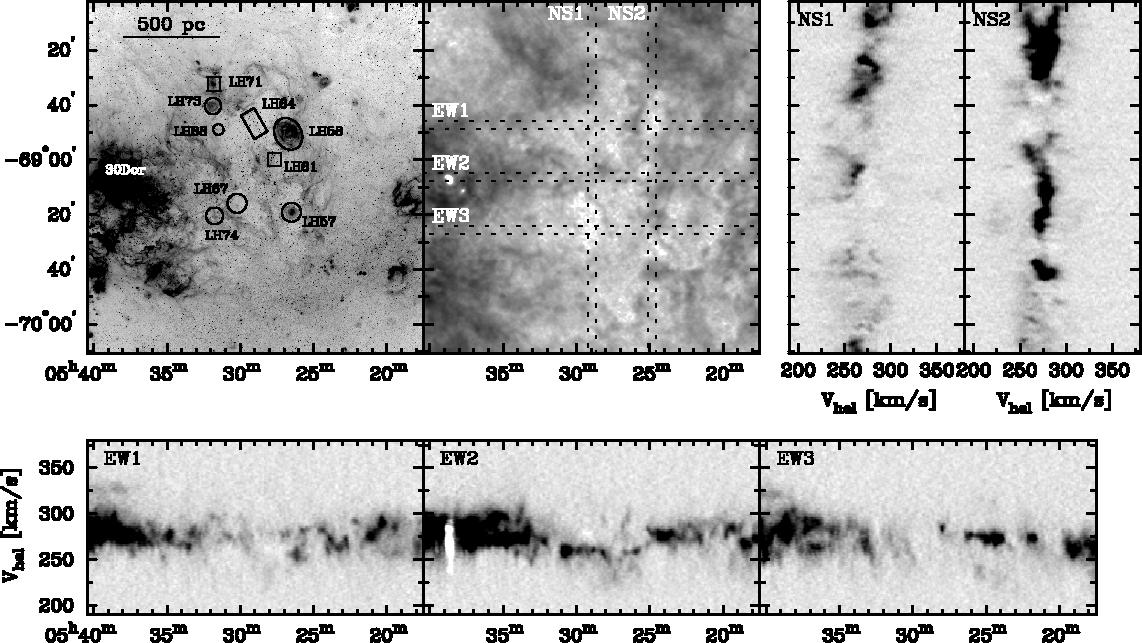}
\caption{(Same as figure \ref{fig:SGS1a}, but Supergiant Shell LMC 3.} 
\label{fig:SGS3a}
\end{figure}

LMC 3 is located just to the northwest of the 30 Doradus complex, 
opposite of LMC 2. It is composed of several curved, 
interconnecting H$\alpha$ filaments, forming a four-lobed shape with 
central coordinates 5$^{\rm h}$30$^{\rm m}$, $-$69$^{\circ}$00$'$ 
(J2000) and a diameter of $\sim$1~kpc (see Fig.~\ref{fig:SGS3a}). LMC 3 has a 
`waist' dividing it into northern and southern lobes, and the region 
across the waist shows significant internal H$\alpha$ and \ion{H}{1} 
structure. The ionized shell lies on the interior of an \ion{H}{1} 
structure along its northern lobes, although the southern ionized 
filaments lie along much less dense regions of neutral gas, which are 
nearly invisible in the \ion{H}{1} column density image. The neutral 
shell extends significantly further to the south than the optical shell.

\begin{figure}[!ht]
\centering
\includegraphics[height=3in]{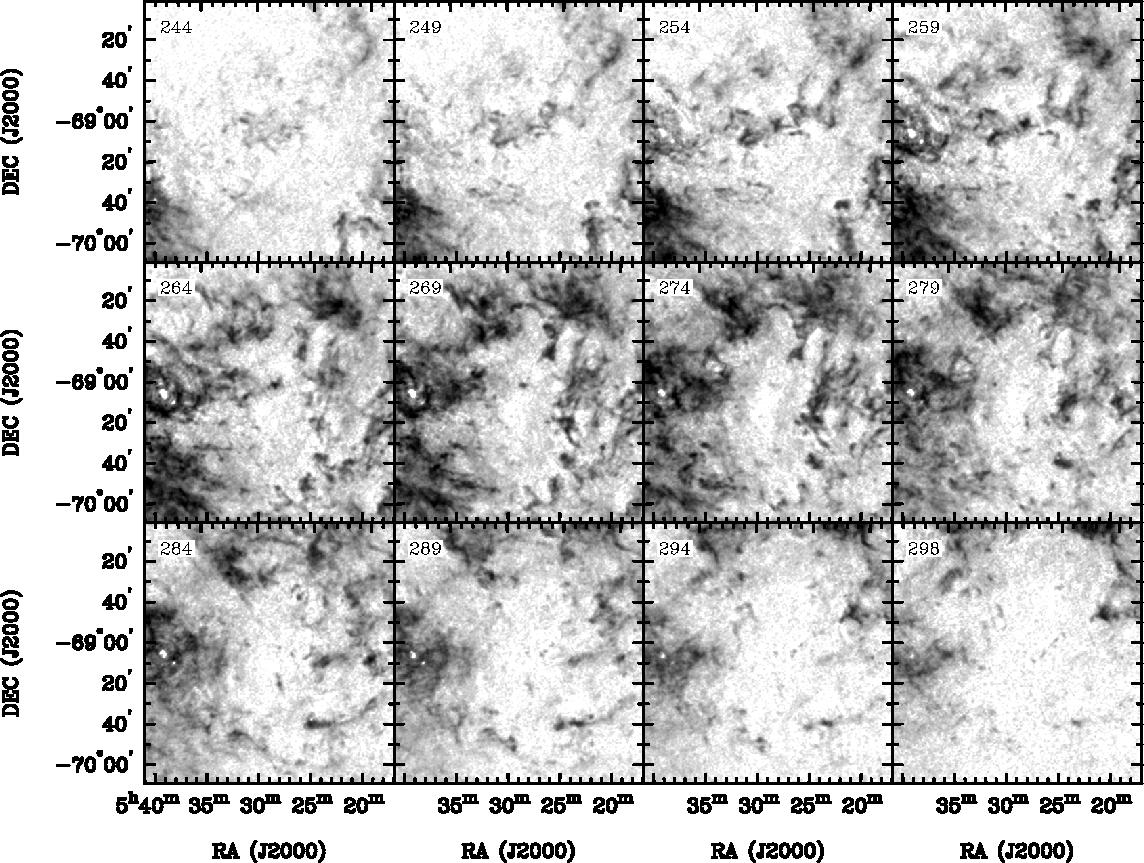}
\caption{(Same as figure \ref{fig:SGS1b}, but Supergiant Shell LMC 3.} 
\label{fig:SGS3b}
\end{figure}

LMC 3 can be detected in channel maps (Fig.~\ref{fig:SGS3b}) in the range 
255~km~s$^{-1}$ to 285 km~s$^{-1}$, with a systemic velocity of 
$\sim$~270~km~s$^{-1}$. Faint \ion{H}{1} structures along the southern 
border of the optical shell can be seen mostly at systemic and 
redshifted velocities. The wall between the two southern lobes, for 
example, appears clearly at 285~km~s$^{-1}$, redshifted  by 15~km~s$^{-1}$ 
with respect to the shell's systemic velocity. The structure along the 
waist of LMC 3 is seen in channel maps to be a broad filament stretching 
from east to west, blueshifted by $\sim$ 15~km~s$^{-1}$ with respect to 
the systemic velocity. A smaller additional shell structure to the west 
of the northern lobes of LMC 3 is clearly visible in channel maps. At 
velocities of 260~km~s$^{-1}$ and less the structure appears distinct 
from LMC 3, but appears connected at higher velocities.

We made two N-S and three E-W P-V diagrams passing through LMC 3. The 
E-W P-V diagram along the waist indicates blueshifted expansion at a 
rate of 15~km~s$^{-1}$ (see Fig.~\ref{fig:SGS3a} EW2), corresponding to 
the blueshifted filament noted from the channel maps. This cut also 
shows significant highly blueshifted material, with velocities as low 
as 210~km~s$^{-1}$, i.e., with an expansion velocity of up to 
60~km~s$^{-1}$ relative to the systemic velocity. This material is also 
observed in N-S P-V diagrams (see Fig.~\ref{fig:SGS3a} NS1), and can be 
identified with the 220 to 230~km~s$^{-1}$ interstellar absorption 
components seen in the UV spectra of stars within LMC 3 \citep{Chu94}. 
The E-W cut through the northern lobes shows evidence of expansion 
(see Fig.~\ref{fig:SGS3a} EW1), while the cut through the southern lobes 
shows only tenuous indications of expansion (see Fig.~\ref{fig:SGS3a} EW3). 
The northern P-V diagram does, however, show possible expansion of the 
smaller western shell, also at a rate of 15~km~s$^{-1}$ but with a 
lower systematic velocity of 260~km~s$^{-1}$. In this image the smaller 
shell appears unconnected to LMC 3. The N-S P-V diagrams also show 
evidence of two-sided expansion for the northern lobes 
(see Fig.~\ref{fig:SGS3a} NS1) and  the smaller western shell 
(see Fig.~\ref{fig:SGS3a} NS2). 

We conclude that LMC 3 is a physical shell with irregular shape and 
expansion pattern, whose \ion{H}{1} shell extends significantly 
beyond the optical shell toward the south. The OB associations show a 
correlation between their ages and locations in the LMC 3 region. Older 
OB associations, that have already blown away their surrounding gas, 
are located in the interior of the shell and may be responsible for its 
irregular expansion pattern: LH67 in the southeast, LH61 along the waist, 
LH64 in the northwest, and LH68 in the northeast. Younger OB associations, 
that are still within dense \ion{H}{2} regions, lie along the rim of 
the optical shell: LH58 near the waist on the western side, LH73 and 
LH71 in the northeast, and LH57 in the southwest. The exception to this 
rule is the older association LH74, which is located along the 
southeastern rim of LMC 3. We find LMC 3 to be the ionized interior 
of the \ion{H}{1} shell SGS 12, whose greater southern extent may be 
explained as the result of the combined supernova explosions and fast 
stellar winds from the older OB associations LH74 and LH67 in the 
southeast of the optical shell. The northern lobes, which are 
significantly brighter in \ion{H}{1}, are identified as a separate 
shell SGS 13, but we conclude that the northern and southern lobes are 
connected to form a single, while irregular, shell.  Further, the 
smaller western shell we identified in channel maps and P-V diagrams 
is associated with SGS 8, although it has no optical counterpart.

\subsection{LMC 4}  \label{sec:lmc4}

LMC 4 is the largest SGS in the LMC, and is located on the northern edge 
of the galaxy. It is roughly circular in shape,  
with central coordinates 5$^{\rm h}$32$^{\rm m}$, $-$66$^{\circ}$40$'$ 
(J2000) and a diameter of 1 kpc (see Fig.~\ref{fig:SGS45a}). One notable 
feature of LMC 4 is a filament of neutral and ionized hydrogen 
extending into the interior from the north and terminating in the OB 
association LH72. The optical shell LMC 4 lies along the interior of 
a larger \ion{H}{1} shell. The density of the neutral shell structure 
is greater than that of the surrounding regions. The densest part of the 
neutral shell is the region to the northwest which forms the border 
between LMC 4 and 5. 

\begin{figure}[!ht]
\centering
\includegraphics[height=3in]{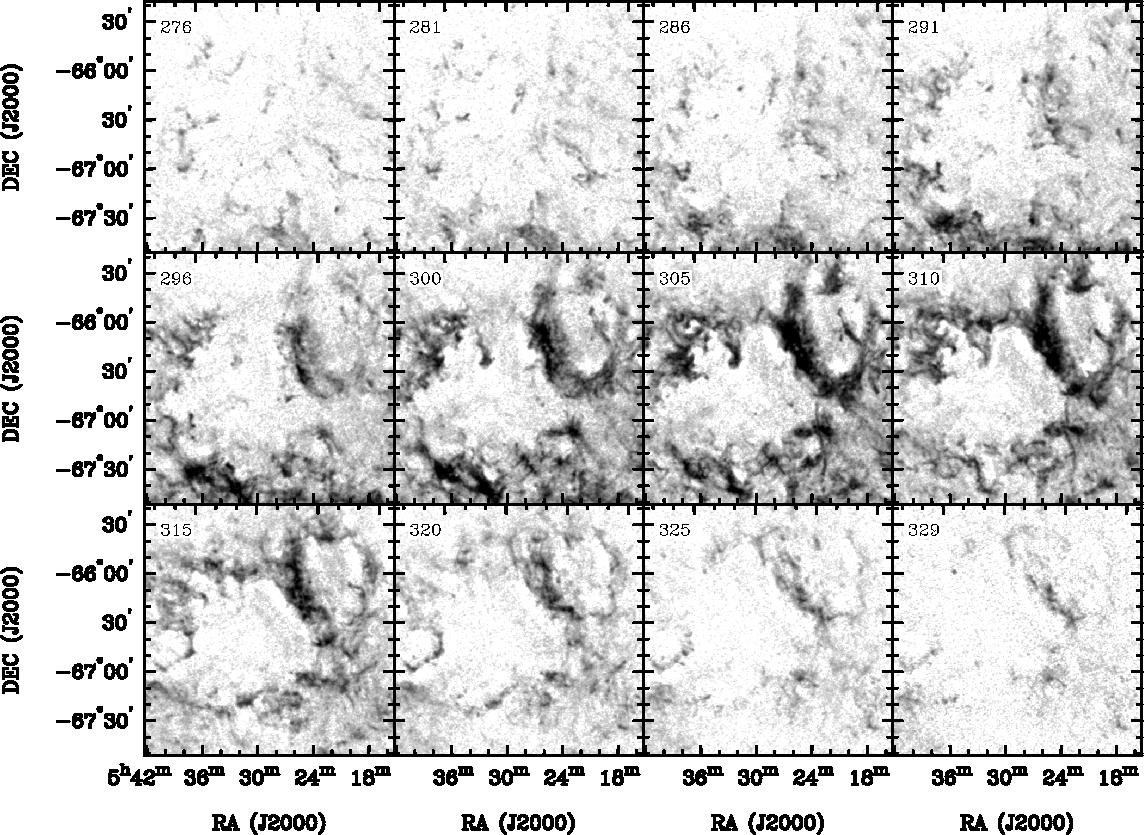}
\caption{Same as figure \ref {fig:SGS1b}, but Supergiant Shells LMC 4 and 5.}
\label{fig:SGS45b}
\end{figure}

LMC 4 can be seen in channel maps (Fig.~\ref{fig:SGS45b}) in the range 
285~km~s$^{-1}$ to 325~km~s$^{-1}$, with a systemic velocity of 
$\sim$~305~km~s$^{-1}$. The southern and eastern edges of the shell are 
blueshifted $\sim$ 10~km~s$^{-1}$ with respect to the systematic velocity, 
while the northern and western edges are redshifted by a similar amount. 
The northwestern corner is the interaction region with LMC 5, and is 
visible in a wide velocity range, also slightly redshifted with respect 
to the range in which the rest of LMC 4 can be seen. We have made three E-W 
and one N-S P-V diagram passing through LMC 4 (see Fig.~\ref{fig:SGS45a} EW1,
EW2, EW3 and NS1). None of these diagrams show evidence of uniform expansion. 
The shell cavity is mostly evacuated, with the interior material present 
in small clouds with velocities blueshifted by up to 40~km~s$^{-1}$ with 
respect to the systemic velocity.

We observe that LMC 4 is a physical shell due to its central cavity and
raised density in the rim. LMC 4 is the ionized inner wall of the 
\ion{H}{1} identified SGS 11. The northeastern corner of this structure 
has been identified as SGS 14; however, the P-V diagram through this 
structure shows clouds only at the systemic velocity, and channel maps 
show that this structure is part of SGS 11. Within the central cavity, 
the most prominent OB association is LH77, also known as Shapley 
Constellation III, a large arc of stars in the southern shell interior. 
This association has an older stellar population, as it has already 
expelled its ambient gas, and has most likely contributed to the formation 
of LMC 4. Along the periphery of LMC 4 are several young OB associations 
still within \ion{H}{2} regions. These include LH91, LH95, and LH83 in 
the northeast, LH52 and LH53 in the northwest along the interaction 
region with LMC 5, and the numerous OB associations in the \ion{H}{2} 
complexes N51, N56, N57 and N59 extending southward from the southern 
rim of LMC 4. The age structure of these OB associations has been 
interpreted as triggered star formation \citep{E98}.

\subsection{LMC 5}  \label{sec:lmc5}

\begin{figure}[!ht]
\centering
\includegraphics[height=3in]{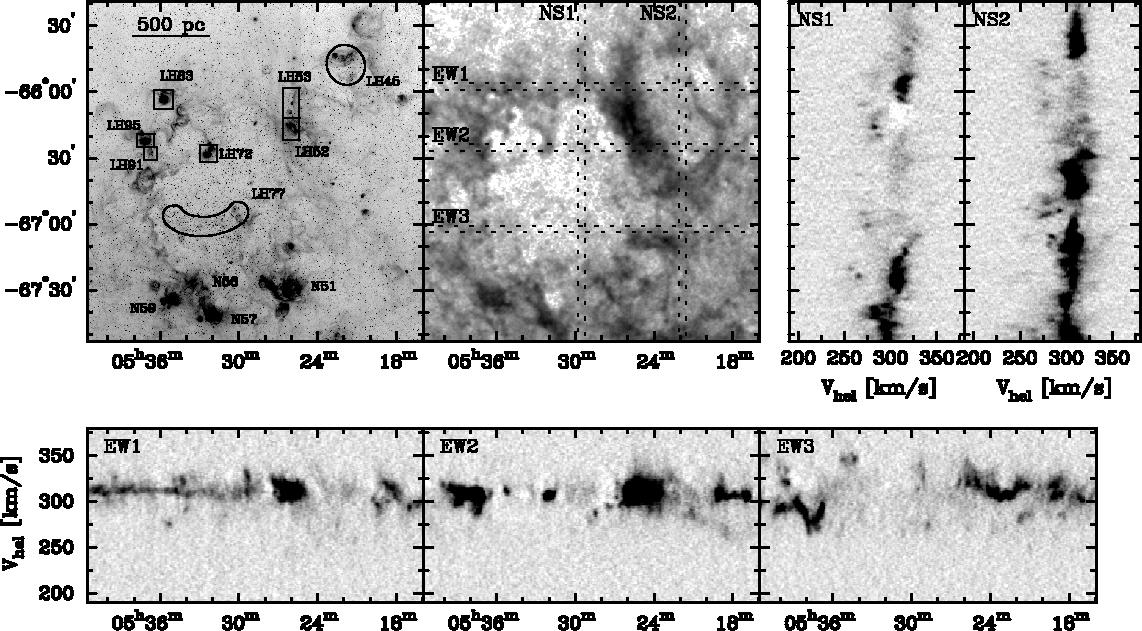}
\caption{Same as figure \ref {fig:SGS1a}, but Supergiant Shells LMC 4 and 5.}
\label{fig:SGS45a}
\end{figure}

LMC 5 has a diameter of 800 pc, and is located directly to the northwest 
of LMC 4 in the north of the LMC with central coordinates 
5$^{\rm h}$22$^{\rm m}$, $-$66$^{\circ}$00$'$ (J2000)
(see Fig.~\ref{fig:SGS45a}). H$\alpha$ filaments delineate LMC 5, with 
brighter filaments in the northwest and east as compared with the 
dimmer northeast and southwest optical filaments. Two distinct H$\alpha$ 
filaments are observed interior to the shell. As in LMC 4, a shell of 
\ion{H}{1} with nearly the same shape as the optical shell surrounds 
LMC 5, with a higher density in the neutral shell than in the 
surrounding regions. The densest part of the neutral shell around LMC 
5 is the interaction region with LMC 4. Both interior filaments apparent 
in H$\alpha$ are also present in the \ion{H}{1} image.

LMC 5 can be seen in channel maps (Fig.~\ref{fig:SGS45b}) in the range 
290~km~s$^{-1}$ to 325~km~s$^{-1}$, with a systemic velocity of 
$\sim$~305~km~s$^{-1}$. There is no significant velocity gradient over 
the \ion{H}{1} shell, and the interior structures are mostly seen around 
the shell's systemic velocity.

We have made one E-W and one N-S P-V diagram passing through LMC 5 
(see Fig.~\ref{fig:SGS45a} EW1 and NS2), from which we find possible 
evidence of expansion (see Fig.~\ref{fig:SGS45a}). In the N-S P-V diagram, 
several blueshifted clouds are visible along the shell, which appear to 
connect to each other and to converge back to the systematic velocity at 
the southern edge of the shell. The E-W P-V diagram shows a similar 
convergence at the western edge, and indicates blueshifted expansion 
with a velocity of at least 50~km~s$^{-1}$. The receding side of LMC 5, 
however, is expanding at only 10~--~20~km~s$^{-1}$.

We conclude that LMC 5 is a physical expanding shell, although its 
approaching and receding sides are quite faint. We have determined that 
LMC 5 is the illuminated interior of \ion{H}{1}-identified SGS 7. The 
OB associations LH45 in the northwest corner and LH52 and LH53 along 
the eastern edge explain the relative brightness of these areas in 
H$\alpha$ as compared with the other sides. It has been suggested that 
the formation of LH52 and LH53 was caused by the compression of gas 
between LMC 4 and LMC 5 \citep{C03}. 

\subsection{LMC 6}  \label{sec:lmc6}

\begin{figure}[!ht]
\centering
\includegraphics[height=3in]{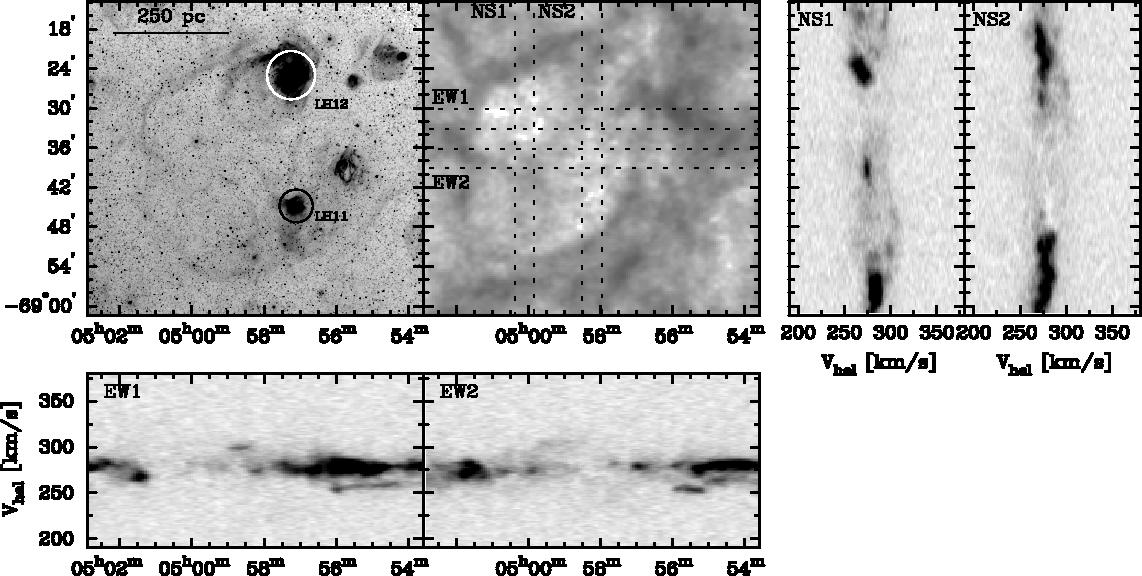}
\caption{Same as figure \ref{fig:SGS1a}, but Supergiant Shell LMC 6.} 
\label{fig:SGS6a}
\end{figure}

LMC 6 is a small, oblong shell with a radius of 600 pc in H$\alpha$, 
located in the southwest of the LMC with central coordinates 
4$^{\rm h}$59$^{\rm m}$, $-$68$^{\circ}$36$'$ (J2000)(see Fig.~\ref{fig:SGS6a}). 
The OB association LH12, in the northeast corner of LMC 6, is in a bright, 
large \ion{H}{2} region N91, from which tails of ionized gas arc to the east. 
The OB association LH11 is located south of LH12 and just below the 
southern edge of LMC 6. The shell is delineated by long ionized filaments 
stretching along its circumference. \cite{M80} considered only the 
ionized filaments, and not LH12, to make up LMC 6, and thus showed it as an 
ellipse open to the northwest. These H$\alpha$ filaments lie along the 
interior of a fairly simple and well defined \ion{H}{1} shell. The neutral 
shell has a dense rim, while the region of LH12 corresponds to an \ion{H}{1} 
hole. There is some \ion{H}{1} structure interior to the shell, running 
northeast to the southwest and from the east to the center. 

\begin{figure}[!ht]
\centering
\includegraphics[height=3in]{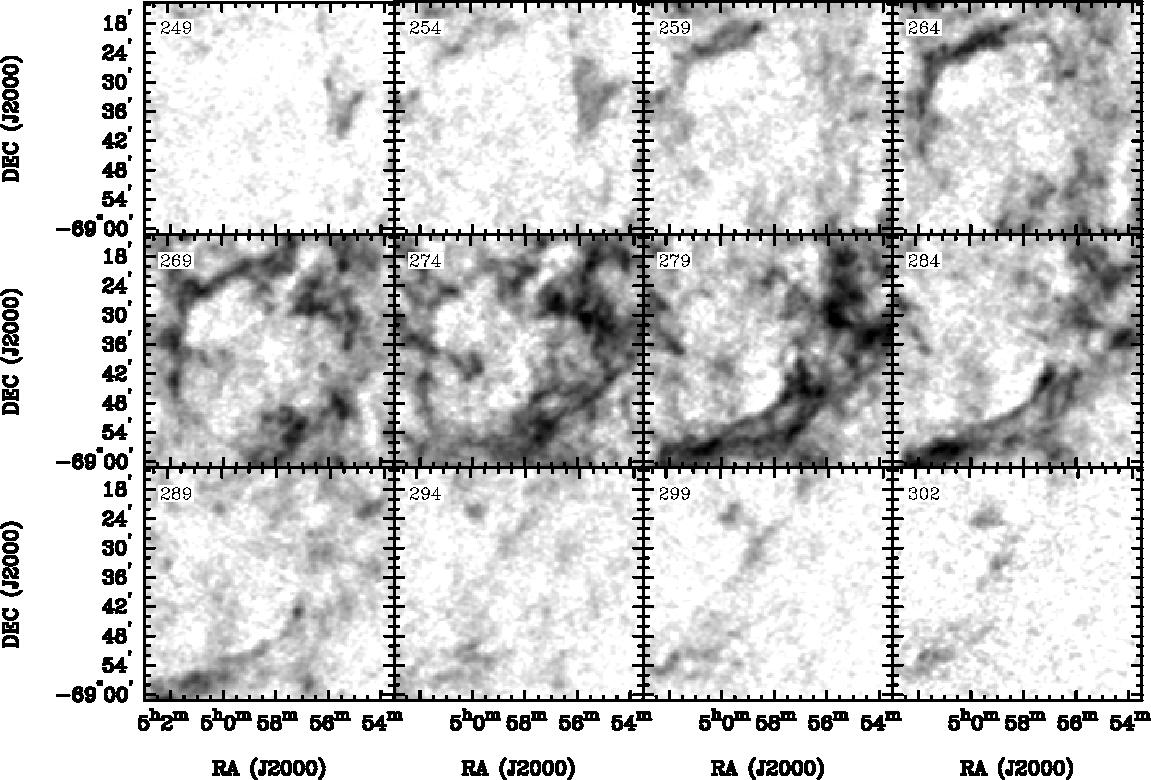}
\caption{Same as figure \ref{fig:SGS1b}, but Supergiant Shell LMC 6.} 
\label{fig:SGS6b}
\end{figure}

LMC 6 is detected in channel maps (Fig.~\ref{fig:SGS6b}) in the range 
260~km~s$^{-1}$ to 290~km~s$^{-1}$, with a systematic velocity of 
$\sim$~275~km~s$^{-1}$. The northeast shell rim is blueshifted with respect 
to the systematic velocity by $\sim$~10~km~s$^{-1}$, and the southeast shell 
rim is redshifted by a similar amount. The channel maps show that the 
\ion{H}{1} structure interior to LMC 6 consists of two distinct filaments, one 
extending to the center of the shell from the eastern wall at the systemic 
velocity, and one stretching from northwest to southeast and redshifted 
$\sim$~25~km~s$^{-1}$ with respect to the systemic velocity.
 
We have made two N-S and two E-W P-V diagrams passing through LMC 6 
(see Fig.~\ref{fig:SGS6a}). The E-W diagrams both show possible expansion, 
with the receding side corresponding to the highly redshifted filament 
observed in the channel maps. We interpret this structure to indicate 
expansion rather than the chance superposition of this high-velocity 
filament over LMC 6 because the high-velocity gas appears to converge to 
the lower velocity gas at the rims of the shell. To the west of LMC 6, a 
region of blueshifted gas is observed which may or may not be connected to 
the rim of the shell. The N-S P-V diagrams show similar indications of 
expansion, while like the E-W diagrams they show that both approaching and 
receding sides of LMC 6 are quite rarefied. We measure typical expansion 
velocities of 20~km~s$^{-1}$, with some material moving at up to 
40~km~s$^{-1}$ with respect to the systemic velocity. The eastern N-S P-V 
diagram (see Fig.~\ref{fig:SGS6a} NS1) shows the dense filament at the systemic 
velocity within LMC 6.

These findings lead us to believe that LMC 6 is a physical, expanding shell, 
and that the ionized shell is the illuminated interior of a larger \ion{H}{1} 
shell corresponding to the \ion{H}{1}-identified SGS 2. The neutral shell 
appears to have an irregular two-sided expansion with a rate ranging between 
20 and 40~km~s$^{-1}$.

\subsection{LMC 7}  \label{sec:lmc7}

\begin{figure}[!ht]
\centering
\includegraphics[height=3in]{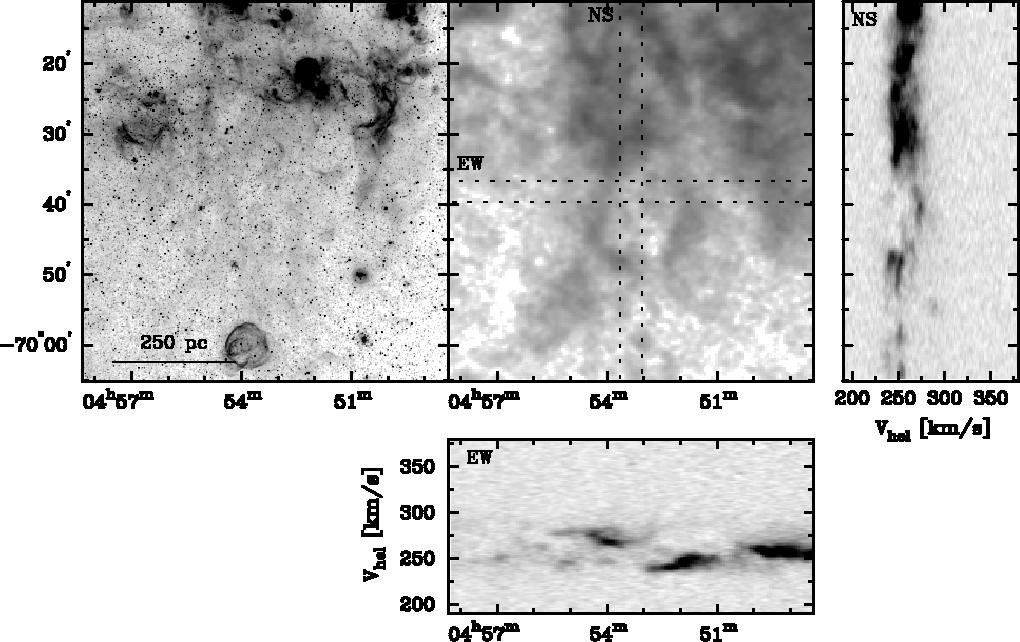}
\caption{Same as figure \ref{fig:SGS1a}, but Supergiant Shell LMC 7.} 
\label{fig:SGS7a}
\end{figure}

LMC 7 is suggested by \cite{M80} to be a possible shell open to the south 
with diameter 800 pc and center at 4$^{\rm h}$53$^{\rm m}$, 
$-$69$^{\circ}$35$'$ (J2000) (see Fig.~\ref{fig:SGS7a}). The H$\alpha$ image 
of this region shows a dense ridge of ionized gas along the north,
with faint interior structure which was interpreted to comprise a shell. 
The interior of LMC 7 contains many filaments and clouds of diffuse 
emission. Inspection of the \ion{H}{1} image reveals little evidence 
for a shell corresponding to LMC 7. The northern rim of ionized gas
corresponds to a similarly dense region of \ion{H}{1}, but also apparent 
is a ridge of \ion{H}{1} extending from the north through the center of 
LMC 7. Neutral hydrogen data does, however, indicate the existence of 
a 225 pc diameter shell located directly south of SGS 7 with center at 
4$^{\rm h}$53$^{\rm m}$, $-$69$^{\circ}$45$'$ (J2000). There is no 
H$\alpha$ emission associated with this structure.

\begin{figure}[!ht]
\centering
\includegraphics[height=3in]{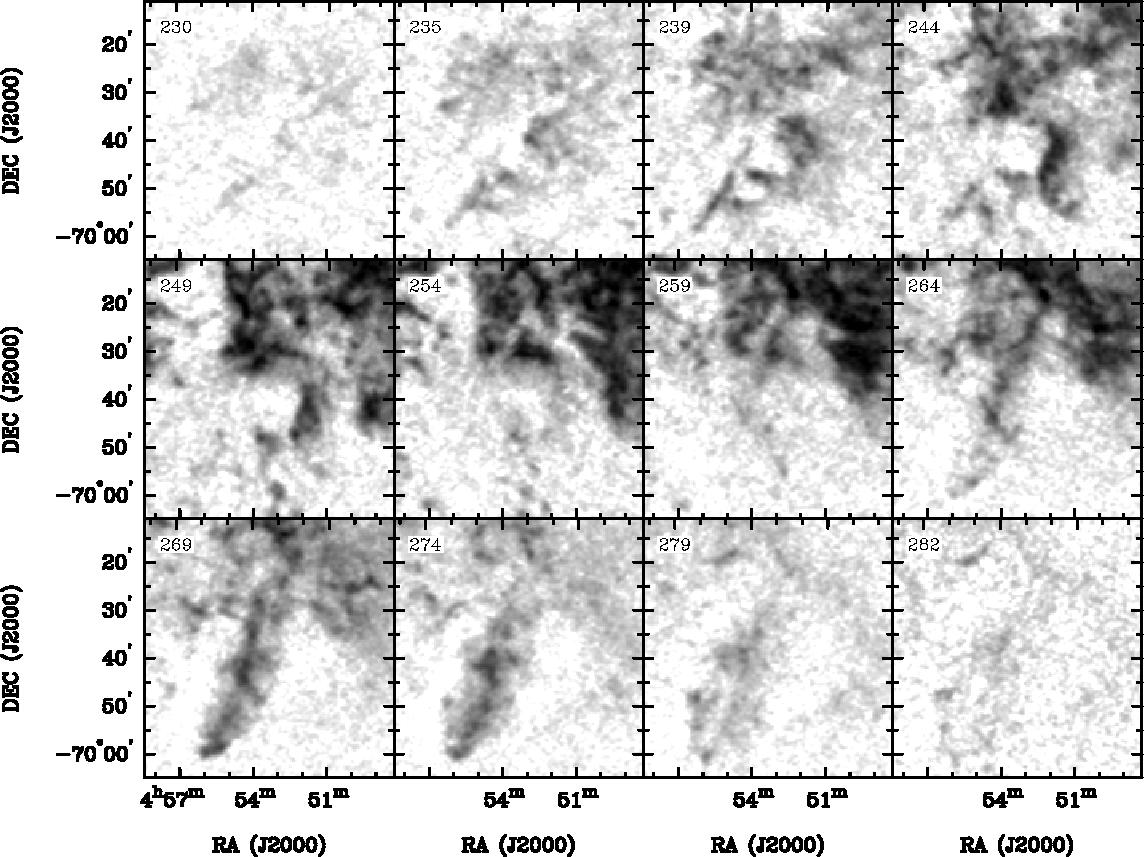}
\caption{Same as figure \ref{fig:SGS1b}, but Supergiant Shell LMC 7.} 
\label{fig:SGS7b}
\end{figure}

Channel maps (Fig.~\ref{fig:SGS7b}) show no shell walls and central cavity for 
LMC 7. The smaller southern shell is shown to have a systemic velocity of 
245~km~s$^{-1}$, and can be seen in the range 235~km~s$^{-1}$ to 
260~km~s$^{-1}$. The ridge of gas passing through the center of LMC 7 is 
seen to be a redshifted structure at \vhel=275~km~s$^{-1}$. N-S and E-W 
P-V diagrams passing through the center of the smaller shell show an 
expansion velocity of 20~km~s$^{-1}$ from the bright emission, and up to 
40~km~s$^{-1}$ from the faint emission on the receding side 
(see Fig.~\ref{fig:SGS7a}). An expansion velocity of 30~km~s$^{-1}$ is 
measured on the approaching side.

We conclude that LMC 7 is not a physical shell due to lack of visible 
\ion{H}{1} shell structure or expansion. However, we note that a smaller 
\ion{H}{1} shell with no optical counterpart and clear signs of expansion 
is located just to the south of LMC 7. Neither LMC 7 nor the smaller 
southern shell was identified by \citet{Kim99}, although 3-4 small
\ion{H}{1} shells along the northern rim were cataloged.

\subsection{LMC 8}  \label{sec:sgs8}

\begin{figure}[!ht]
\centering
\includegraphics[height=3in]{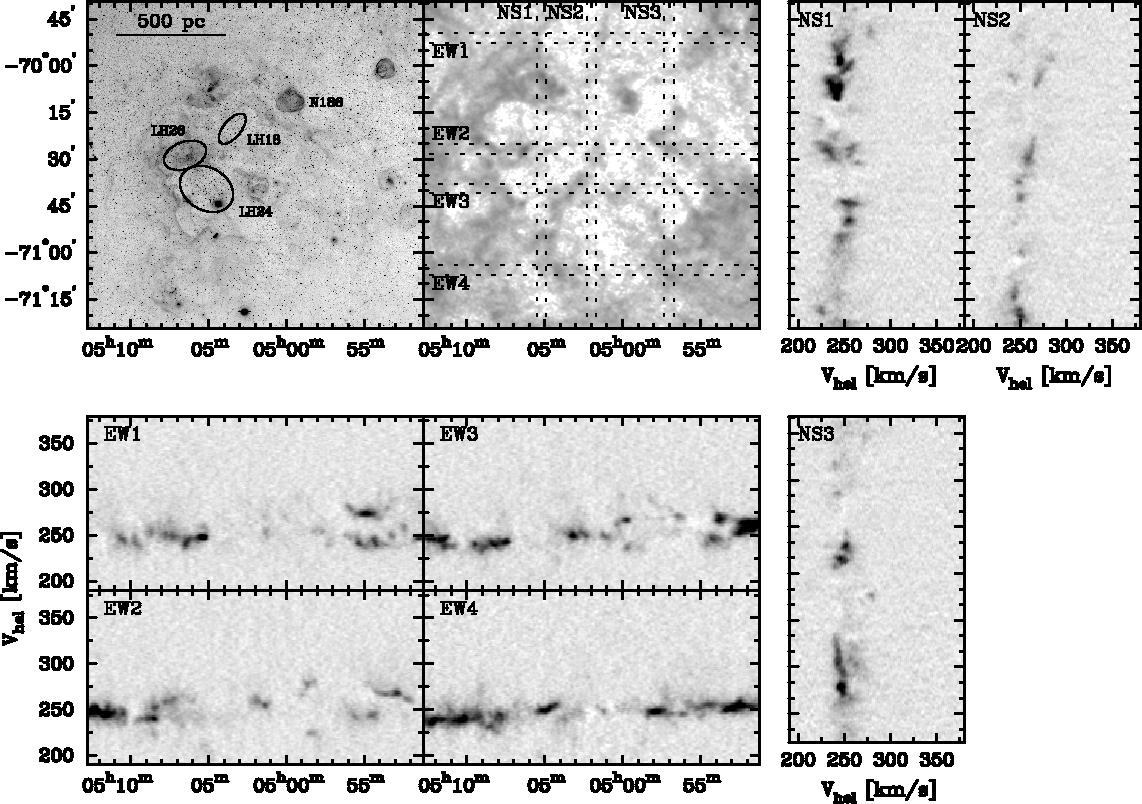}
\caption{Same as figure \ref{fig:SGS1a}, but Supergiant Shell LMC 8.} 
\label{fig:SGS8a}
\end{figure}

LMC 8 is a 900 pc diameter shell open to the southwest and located in the 
southwestern corner of the LMC, with central coordinates 
5$^{\rm h}$03$^{\rm m}$, $-$70$^{\circ}$30$'$ (J2000) 
(see Fig.~\ref{fig:SGS8a}). The H$\alpha$ structure is by no means a clean 
shell, consisting of filaments to the southeast and northwest, with 
significant amounts of internal diffuse emission in the east of the shell. 
The \ion{H}{1} structure is significantly more complex, with a 
multi-lobed, `flower' shape. The eastern lobe appears to have H$\alpha$ 
emission interior to the neutral hydrogen, while the other lobes do not
have obvious optical counterparts. Some \ion{H}{1} filaments and clouds 
have optical counterparts; for example, an isolated spot of \ion{H}{1} 
between the northern and northwestern lobes corresponds to the superbubble 
N186 \citep[=DEM L50;][]{R90,O02}.

\begin{figure}[!ht]
\centering
\includegraphics[height=3in]{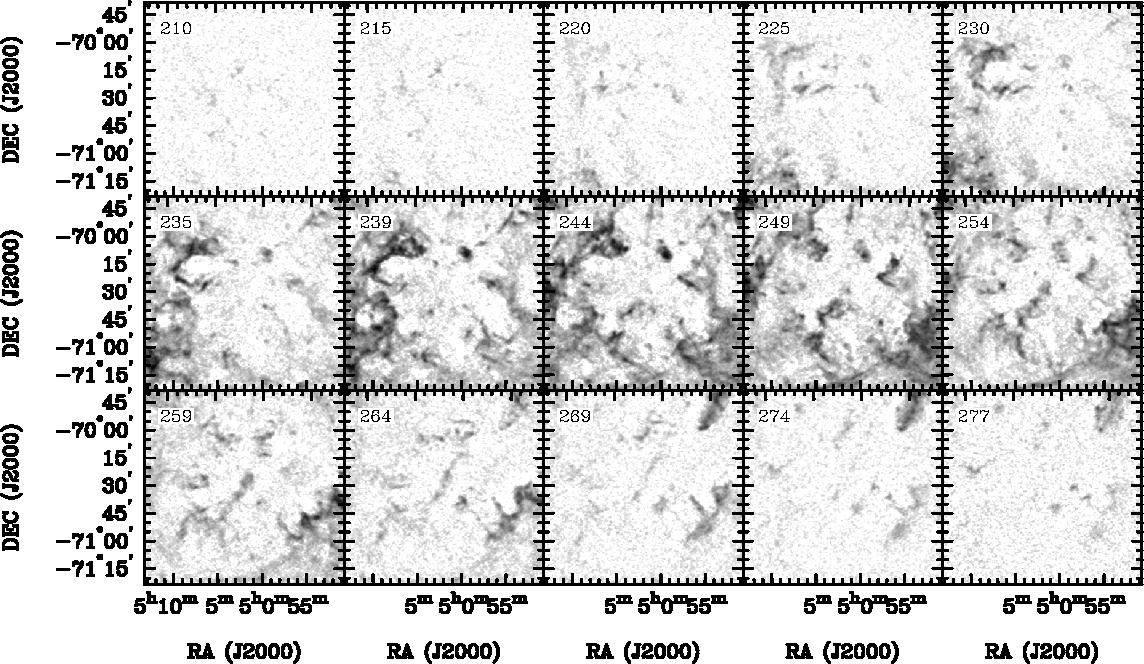}
\caption{Same as figure \ref{fig:SGS1b}, but Supergiant Shell LMC 8.} 
\label{fig:SGS8b}
\end{figure}

LMC 8 is detected in \ion{H}{1} channel maps (Fig.~\ref{fig:SGS8b}) between 
225~km~s$^{-1}$ and 270~km~s$^{-1}$. The systemic velocity of the overall 
structure is 245~km~s$^{-1}$. The lobes of LMC 8 appear to be several 
interlocking shells with somewhat different systemic velocities. The 
northern lobe has a systemic velocity of 250~km~s$^{-1}$ and center 
located at 5$^{\rm h}$03$^{\rm m}$, $-$69$^{\circ}$50$'$ (J2000). The 
northeastern and eastern lobes have systemic velocities of 240~km~s$^{-1}$ 
and 250~km~s$^{-1}$, and centers located at 
5$^{\rm h}$06$^{\rm m}$, $-$70$^{\circ}$20$'$ (J2000) and 
5$^{\rm h}$06$^{\rm m}$, $-$70$^{\circ}$35$'$ (J2000), respectively. The 
southern and southwestern lobes have systemic velocities of 250~km~s$^{-1}$ 
and 240~km~s$^{-1}$, and centers located at 
5$^{\rm h}$02$^{\rm m}$, $-$71$^{\circ}$00$'$ (J2000) and 
4$^{\rm h}$58$^{\rm m}$, $-$70$^{\circ}$30$'$ (J2000), respectively. The 
channel maps show that the interiors of these separately expanding lobes are 
connected at 245~km~s$^{-1}$.

We have made 3 N-S and 4 E-W P-V diagrams to survey the expansion of the 
lobes of LMC 8. The cut NS1 (see Fig.~\ref{fig:SGS8a}) reveals 
that the eastern and northeastern lobes are expanding at a rate of up to 
35~km~s$^{-1}$, and NS2 shows that the northern lobes are expanding at up 
to 30~km~s$^{-1}$ while the southern lobes have no distinct expansion. NS3   
shows possible one-sided, redshifted expansion at a rate of up to 
30~km~s$^{-1}$ for the southwestern lobe. 

The E-W P-V diagrams give further information about the expansion structure 
of LMC 8. They reveal that the northern lobe extends much further to the west 
than is apparent in the column density image (see Fig.~\ref{fig:SGS8a} EW1). The 
northern center of LMC 8 shows highly blueshifted material down to 
200~km~s$^{-1}$ (see Fig.~\ref{fig:SGS8a} EW2). The western lobes show a 
complicated expansion pattern with at least two distinct systemic velocities
(see Fig.~\ref{fig:SGS8a} EW3 and EW4). 

We conclude that LMC 8 is comprised of several physical, expanding shells 
that are connected. The optical shell is associated with the 
\ion{H}{1}-identified SGS 4, but covers only the eastern portion of the
neutral shell.  The shape of the optical shell may be explained by the 
positions of the OB associations LH26, LH18 and LH24 in the north and east
of LMC 8, thus illuminating only the northern and eastern walls of the neutral 
shell and creating the optical appearance of a shell open to the southwest.  

\subsection{LMC 9}  \label{sec:lmc9}

\begin{figure}[!ht]
\centering
\includegraphics[height=3in]{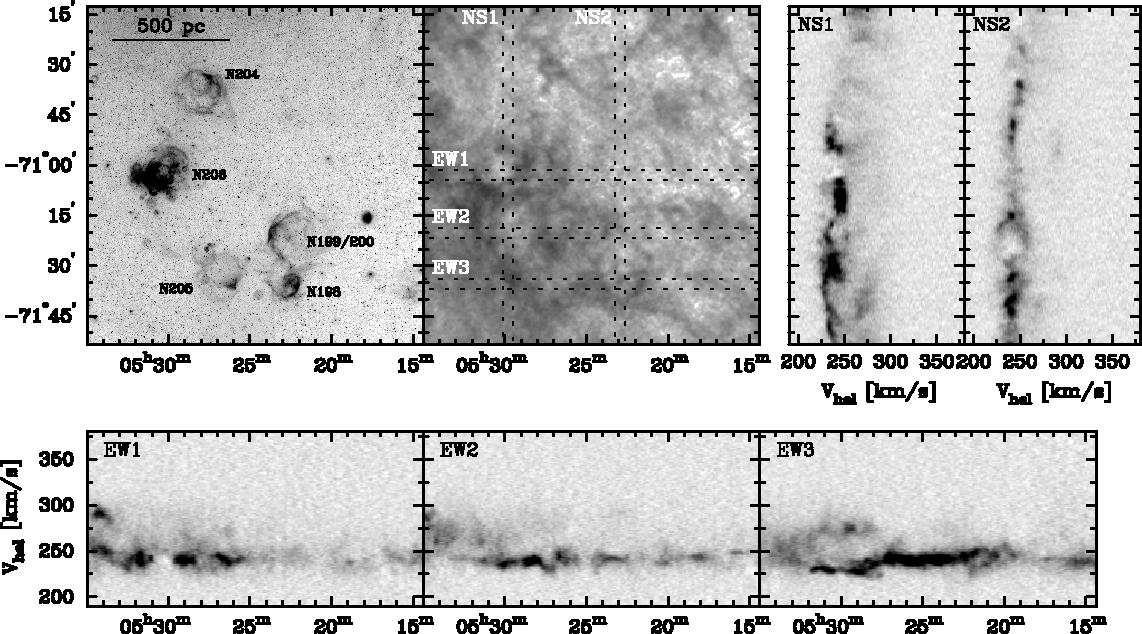}
\caption{Same as figure \ref{fig:SGS1a}, but Supergiant Shell LMC 9.}
\label{fig:SGS9a}
\end{figure}

The optical shell LMC 9 is located in the southern part of the LMC, with
central coordinates 5$^{\rm h}$25$^{\rm m}$, $-$71$^{\circ}$05$'$ (J2000) 
and diameter 890 pc (see Fig.~\ref{fig:SGS9a}). It is a collection of supershells 
situated in a semi-circle: N199/200, N198, N205, N206 and N204, from west to 
northeast along the periphery, connected with faint filaments and open to the
northwest \citep{M80}. The \ion{H}{1} data shows generally higher column 
density toward the southeast quadrant of LMC 9. The supershells and their 
connecting filaments along LMC 9 have no corresponding \ion{H}{1} shell rim, 
but rather \ion{H}{1} holes near each of the supershells. 

\begin{figure}[!ht]
\centering
\includegraphics[height=3in]{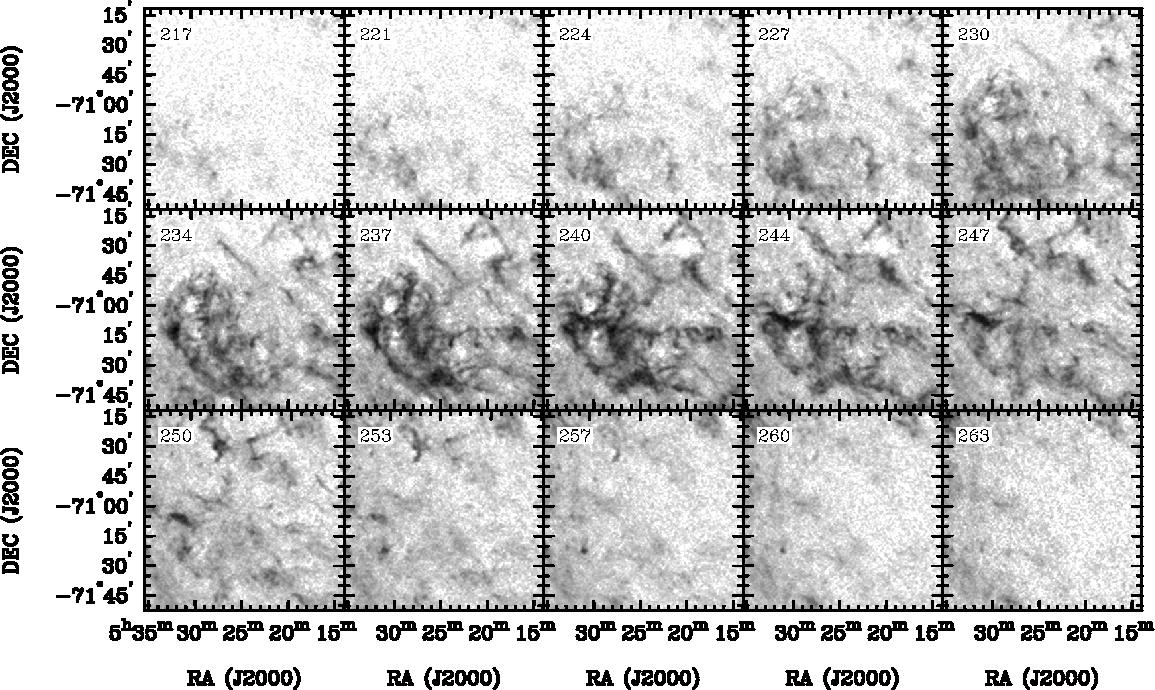}
\caption{Same as figure \ref{fig:SGS1b}, but Supergiant Shell LMC 9.}
\label{fig:SGS9b}
\end{figure}

The region of SGS 9 is seen in channel maps (Fig.~\ref{fig:SGS9b}) to be full 
of \ion{H}{1} diffuse emission and filaments, with no distinct central 
cavity. Several smaller cavities can be seen, at least two of which 
correspond to individual supershells: the supershell N206 whose \ion{H}{1} 
cavity is detected in the range 225~km~s$^{-1}$ to 245~km~s$^{-1}$, and the 
supershell N199/200 whose \ion{H}{1} cavity is detected in the range 
225~km~s$^{-1}$ and 250~km~s$^{-1}$. Also visible is an oblong cavity to 
the southeast centered at 5$^{\rm h}$30$^{\rm m}$, $-$71$^{\circ}$20$'$ 
(J2000), which is detected in the range 230~km~s$^{-1}$ to 250~km~s$^{-1}$ 
but has no optical counterpart.

We have made two N-S and three E-W P-V diagrams in the LMC 9 region to 
investigate its expansion structure. The P-V diagrams indicate expansion 
of the southwest supershell (encompassing N198 and N199/200) at a rate 
of up to 20~km~s$^{-1}$ (see Fig.~\ref{fig:SGS9a} EW2, EW3 and NS2), as well 
as possible expansion of the southeast oblong cavity at a rate of up to 
20~km~s$^{-1}$ (see Fig.~\ref{fig:SGS9a} EW2 and NS1). The most clearly 
expanding area in the LMC 9 region, however, is a region in the southeast 
corner of our frame, outside the boundary of LMC 9, where no cavity is 
visible in the H$\alpha$ or \ion{H}{1} column density maps. This expanding 
structure, centered at 5$^{\rm h}$32$^{\rm m}$, $-$71$^{\circ}$35$'$ 
(J2000), has an expansion velocity of up to 40~km~s$^{-1}$ (see 
Fig.~\ref{fig:SGS9a} EW3 and NS1).

While \citet{Kim99} identified the \ion{H}{1} shell SGS 9 as the 
counterpart of LMC 9, we conclude that LMC 9 is not a physical, 
coherent shell since it has no well-defined \ion{H}{1} counterpart or 
expansion pattern. Some of the supershells along the periphery of LMC 9 appear 
to be expanding shells.

\section{Discussion}  \label{sec:disc}

Comparison of neutral and ionized hydrogen images has allowed us to gain an 
understanding of the structure and kinematics of the optically identified 
SGSs. We have determined that these shells can be separated into three 
categories: 

\begin{itemize}
\item {\bf Simple Coherent Shells:} These are neutral shells whose inner 
walls are photoionized. While the stars in the interior of the SGSs are
responsible for the expanding structure, most of the ionizing sources are
distributed along the periphery. Of the optically identified LMC SGSs, 
LMC 1, 4, 5 and 6 fall within this category.

\item {\bf Complex Shells:} These are structures in which the ionized 
shell delineates only part of a larger or more complex neutral shell. The 
complex shells include LMC 2, 3 and 8. In the cases of LMC 2 and LMC 3, 
the optical shell illuminates only the northern portion of an larger 
neutral structure, while the single optically identified shell LMC 8 
appears to contain multiple expanding neutral structures, of which only
the eastern part is associated with ionized gas.

\item {\bf False Shells:} These are those optically identified 
structures whose neutral \ion{H}{1} counterpart does not show
any of the following characteristics to indicate a shell structure:
(1) expansion within the shell boundary, as expected from an 
expanding shell; (2) raised column densities along shell rims at 
the systemic velocity, showing a shell morphology; and (3) central 
cavity at the systemic velocity, or for an \ion{H}{1} hole.
A physical shell, even if it has been stalled, can still be recognized
by the latter two characteristics.
We found no evidence in neutral hydrogen to confirm the shell structure 
of LMC 7 and LMC 9, and characterize them as false shells.  LMC 7 lacked 
any corroborating \ion{H}{1} structure, while the optically identified 
shell LMC 9 was seen to be a chance superposition of ionized filaments 
and OB associations, with a neutral hydrogen structure in no way 
indicating a shell.
\end{itemize} 

The nature of the optically identified SGSs can be understood as the 
result of the interplay between OB associations and the ISM. 
The fast stellar winds and supernova explosions from OB
associations clear away the interstellar gas and form a large expanding
shell structure.  The initial OB associations will lose their ionizing
power after $\sim$12 Myr, the lifetime of a 15 $M_{\odot}$ star, the
least massive star whose UV flux still produces detectable \ion{H}{2}
gas.  The recombination timescale of ionized gas is  
$7.6\times10^4 N_{\rm e}^{-1}$ yr, where $N_{\rm e}$ is the electron 
density in cm$^{-3}$.  The interstellar gas density in a SGS is most
likely in the range of 0.1--1 cm$^{-3}$, and the recombination timescale
is much less than 1 Myr.  Thus, a SGS with no new OB associations will 
cease to be observable in H$\alpha$ soon after 12 Myrs, and only an 
\ion{H}{1} shell will be detected.

A summary of the diameters ($D$), average expansion velocities 
($V_{\rm exp}$), and maximum expansion velocities ($V_{\rm max}$)
of the optically identified SGSs in the LMC is given in Table \ref{tab:SGS}.
We estimate their ages from these data using the formula 
$t \simeq \eta (D/2) V_{\rm exp}^{-1}$, where $\eta$ is 0.4 for a 
momentum-conserving bubble \citep{S75}, and 0.6 for an energy-conserving 
bubble \citep{C75, W77}. We adopt an intermediate value $\eta=0.5$. 
Examining the ages derived in Table \ref{tab:SGS}, we find that all but 
one, SGS 8, are less than 12 Myr old. SGS 8, however, contains the OB 
associations LH18, LH24 and LH26 (see section \ref{sec:sgs8}), of which 
LH24 and LH26 are still young enough to contain \ion{H}{2} gas. Those
that are younger than 12 Myr all have young OB associations along their
peripheries.  These young OB associations explain the presence of an 
optical counterpart to this \ion{H}{1} shell, regardless of their ages.

Of the \ion{H}{1}-identified SGSs \citep{Kim99}, 15 have average 
diameters $\geq$ 600 pc, the threshold used for the optical SGSs:
SGS 1, 3, 4, 5, 6, 7, 9, 10, 11, 12, 17, 18, 19, 20, and 23.
Some of these \ion{H}{1}-identified SGSs are associated with the
H$\alpha$-identified SGSs: SGS 3 with LMC 1, SGS 4 with LMC 8, 
SGS 7 with LMC 5, SGS 9 with LMC 9, SGS 11 with LMC 4, SGS 12 
with LMC 3, and SGS 19 and 20 with LMC 2.  As described in \S 4
and discussed earlier in this section, the H$\alpha$-identified
SGSs may correspond to partial or complete walls of 
\ion{H}{1}-identified SGSs, depending on the distribution of 
OB associations that provide ionizing fluxes.
Among the other \ion{H}{1} SGSs that are not associated with H$\alpha$
shells, SGS 1 and 6 do not contain any OB associations, and hence 
have no ionized counterparts; SGS 5 and 23 have no OB 
associations or \ion{H}{2} regions along the most well-defined
shell rims; SGS 10 and 18 do not show obvious shell structures
in either column density or channel maps, so they may not be physical
shells; SGS 17's northern rim is coincident with a long H$\alpha$ arc
that is likely to be ionized by massive stars in the 30 Doradus complex 
in the south \citep[the H$\alpha$ arc is sketched in dashes in 
Fig.\ 1 of][]{M80}.  These comparisons adequately illustrate the
importance of the presence of young OB associations to photoionize
the gas in \ion{H}{1} SGSs to produce H$\alpha$-identified shell
morphology.

Supergiant shells have been identified in distant galaxies based on 
H$\alpha$ images \citep[e.g.,][]{HG97,M98}. However, from our 
analysis of the LMC H$\alpha$ SGSs, it is clear that the H$\alpha$ 
morphology alone is not sufficient to identify physically expanding 
shells and that the H$\alpha$ SGSs represent only the SGS population
that possess young OB associations. High-resolution \ion{H}{1} 
position-velocity datacubes are needed to verify the shells' structure
and to reveal the entire population of SGSs.

\section{Summary}  \label{sec:sum}

Supergiant shells are the largest interstellar structures in galaxies.
They can puncture the gas disk and vent hot gas into the halo, or may induce
further star formation in the disk as they expand. Therefore, they play a 
very important role in the global structure and evolution of the ISM. We
have chosen to study SGSs in the LMC because of its proximity, nearly face-on 
orientation, and small extinction. 

Nine SGSs with diameters $\geq$ 600 pc have been identified in the LMC based 
on H$\alpha$ images, and twenty-three SGSs with diameters $\geq$ 360 pc have 
been reported based on \ion{H}{1} 21-cm line observations. Fifteen of these
twenty-three have sizes $\geq$ 600 pc, of which eight are associated with
H$\alpha$ identified shells. Clearly, these sets do not always identify the 
same structures.

We have examined the physical structure of the optically identified SGSs using 
MCELS H$\alpha$ images and ATCA+Parkes \ion{H}{1} channel maps and P-V 
diagrams to analyze the gas distribution and kinematics. 

Of the nine optically identified SGSs, seven appear to be true shells, where
four are the ionized inner walls of \ion{H}{1} SGSs and three are an ionized 
portion of a larger and more complex \ion{H}{1} structure. All of the 
H$\alpha$ SGSs have OB association along the periphery or in the center, with 
younger OB associations more often found along the periphery. After roughly 
12 Myrs, if no new OB associations have been formed a SGS will cease to be 
identifiable at visible wavelengths. Thus, to be identified in optical 
wavelengths, a SGS must have recent star formation along its periphery.
Based on our analysis, H$\alpha$ observations alone cannot unambiguously 
identify SGSs, especially in distant galaxies. High resolution \ion{H}{1}
observations are needed to reveal the real population of SGSs. 

%
%

\bibliographystyle{apj}  
\bibliography{ref}

\clearpage

\begin{table}[!ht]
	\caption{Properties of Optically Identified SGSs}
	\label{tab:SGS}
	\begin{center}
		\begin{tabular}{|cccccc|}		\hline
			LMC SGS		&		Center (J200)	&		Diameter [pc]	&		$V_{\rm exp}$ [km s$^{-1}$] 	&
 		$V_{\rm max}$ [km s$^{-1}$] & Age [Myrs]  \\   \hline
			LMC 1 & 5$^{\rm h}$00$^{\rm m}$, $-$65$^{\circ}$40$'$	&	750	 & 15 & 25 (red)  & 11.7	\\
			LMC 2 & 5$^{\rm h}$44$^{\rm m}$, $-$69$^{\circ}$20$'$	&	900	 & 25 & 40        & 9	    \\
			LMC 3 & 5$^{\rm h}$30$^{\rm m}$, $-$69$^{\circ}$00$'$	&	1000 & 25 & 40 (blue) & 10    \\
			LMC 4 & 5$^{\rm h}$32$^{\rm m}$, $-$66$^{\circ}$40$'$	&	1000 & -  & -         & -	    \\	
			LMC 5 & 5$^{\rm h}$22$^{\rm m}$, $-$66$^{\circ}$00$'$	&	800	 & 30 & 50        & 6.7 	\\
			LMC 6 & 4$^{\rm h}$59$^{\rm m}$, $-$68$^{\circ}$36$'$	&	600	 & 20 & 40        & 5   	\\
			LMC 7 & 4$^{\rm h}$54$^{\rm m}$, $-$69$^{\circ}$33$'$	&	800	 & -  &  -        & -	    \\
			LMC 8 & 5$^{\rm h}$04$^{\rm m}$, $-$70$^{\circ}$30$'$	&	900	 & 15 & 40        & 15  	\\
			LMC 9 & 5$^{\rm h}$26$^{\rm m}$, $-$71$^{\circ}$00$'$	&	890	 & -  & -         & -	    \\		\hline
		\end{tabular}
	\end{center}
\end{table}

\clearpage

\end{document}